\documentclass{article}
\usepackage{spconf,amsmath,graphicx,hyperref,booktabs,multirow}
\usepackage{pgfplots}
\usepackage{placeins}
\usepgfplotslibrary{groupplots}
\pgfplotsset{compat=1.18}

\title{Acoustic Progress Propagation for\\
Long-Horizon Speculative Decoding in ASR}
\name{Yuanyuan Jia \qquad Qianqian Yang}
\address{Zhejiang University, Hangzhou, China}
\begin{document}
%
\maketitle
\begin{abstract}
Speculative decoding accelerates autoregressive automatic speech recognition
(ASR), but the acceptance length of alignment-aware drafters can saturate as
the draft horizon increases. We propose a progress-aware speculative drafter
that recurrently propagates an acoustic progress state across draft steps
and feeds it back into audio cross-attention to guide token generation. We jointly
train the drafter and progress predictor over variable draft horizons. On five ASR test
sets, our method achieves lossless, macro-averaged end-to-end speedups of
$1.657\times$ and $1.227\times$ over target-only autoregressive decoding with
Qwen3-ASR-0.6B and Qwen3-ASR-1.7B, respectively. Relative to AnchorDraft, our method improves the macro-averaged speedup
by 34.3\% and 9.0\%, respectively. Horizon sweeps show continued growth
in acceptance length beyond the baselines' saturation.
Code is available at
\url{https://github.com/yuanyuanjia71-spec/ProgDraft}.
\end{abstract}
\begin{keywords}
Automatic speech recognition, speculative decoding, inference acceleration
\end{keywords}
\section{Introduction}
\label{sec:intro}

Recent state-of-the-art automatic speech recognition (ASR) models typically
employ an autoregressive Transformer-based decoder to generate transcriptions
conditioned on acoustic representations~\cite{radford2023whisper,shi2026qwen3asr}.
However, this autoregressive paradigm generates tokens one by one, resulting
in high inference latency that limits its ability to meet the requirements
of latency-sensitive applications, such as real-time transcription and voice
assistants~\cite{wei2025specasr}.
To overcome this limitation, speculative
decoding~\cite{leviathan2023speculative,chen2023speculative,li2024eagle,chen2026dflash}
has emerged as a possible solution. It uses a lightweight draft model to
propose multiple candidate tokens, which the target model verifies in parallel
within a single forward pass, reducing sequential decoding overhead.

Recent works have explored this approach for speech recognition.
Whisper-Medusa~\cite{segalfeldman2024whispermedusa} extends Medusa-style
multi-head prediction~\cite{cai2024medusa} to ASR, while
SpecASR~\cite{wei2025specasr} accelerates decoding
through adaptive draft lengths and draft reuse.
However, generating accurate ASR drafts also requires tracking which speech
segment corresponds to each predicted token. Because different tokens span
different amounts of audio, advancing the text by one word or subword does
not directly show where to look next in the speech. A lightweight drafter
must therefore track its own acoustic position across multiple steps.
Recent work~\cite{wang2026alignmentdrift} also shows that this audio attention
often drifts away from the correct position during rollouts, even when the
drafter can see the full audio context. This drift lowers token acceptance
at later steps, limiting the speedup of longer drafts.
To address this problem, AnchorDraft adds stepwise supervision to the
drafter's audio attention during training~\cite{wang2026alignmentdrift}.

However, improving per-step alignment accuracy does not necessarily enable
effective use of longer drafts. We observe pronounced long-horizon saturation.
As draft length increases, AnchorDraft's acceptance length quickly
saturates, and additional draft steps yield little further decoding progress.
Stepwise alignment supervision can improve the drafter's acoustic localization
accuracy, but it does not change how localization is performed at inference
time. At each draft step, the drafter must use its current state to re-identify
the acoustic region corresponding to the next token within the full acoustic
context~\cite{wang2026alignmentdrift}. Thus, accurate local alignment over short horizons does not ensure
that the drafter can continuously and reliably track its position in the audio
as the draft trajectory advances.

Building on prior recurrent position updates and Gaussian acoustic-attention
weighting~\cite{tjandra2017local}, we propose a progress-aware speculative
drafter that propagates acoustic progress from a target-attention anchor
across draft steps to improve long-horizon acceptance.
We further adopt variable-horizon training, randomly sampling draft lengths
during training so that the model learns this acoustic progression over
rollouts of different lengths, thereby improving stability and enabling more
effective use of longer draft horizons.

Our contributions are threefold:
\begin{enumerate}
\item We propose acoustic progress propagation to guide successive speculative
ASR draft steps through recurrent updates of an acoustic position state
and feedback to audio cross-attention.
\item We jointly train the drafter
and progress predictor over variable draft horizons, combining token
prediction and alignment supervision to learn acoustic progress transitions
across rollouts of different lengths.
\item Experiments on five ASR test sets show that our method makes more
effective use of long draft horizons, achieving longer acceptance lengths
and higher lossless end-to-end speedups than the evaluated baselines.
\end{enumerate}

\begin{figure}[t]
\centering
\begin{minipage}{\columnwidth}
\centering
\includegraphics[width=\linewidth]{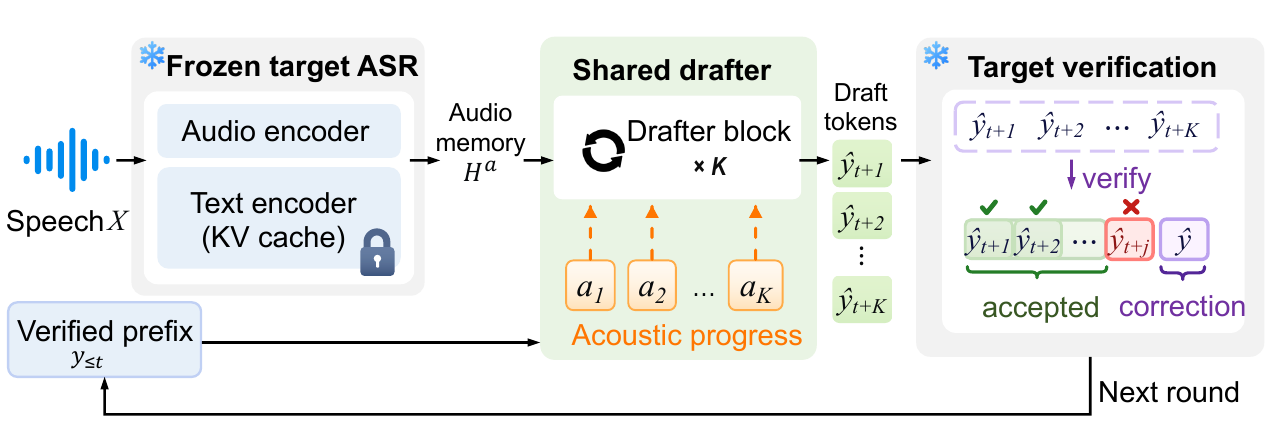}
\par\smallskip
{\small (a) Progress-conditioned speculative ASR decoding\par}
\end{minipage}
\par\medskip
\begin{minipage}{\columnwidth}
\centering
\includegraphics[width=\linewidth]{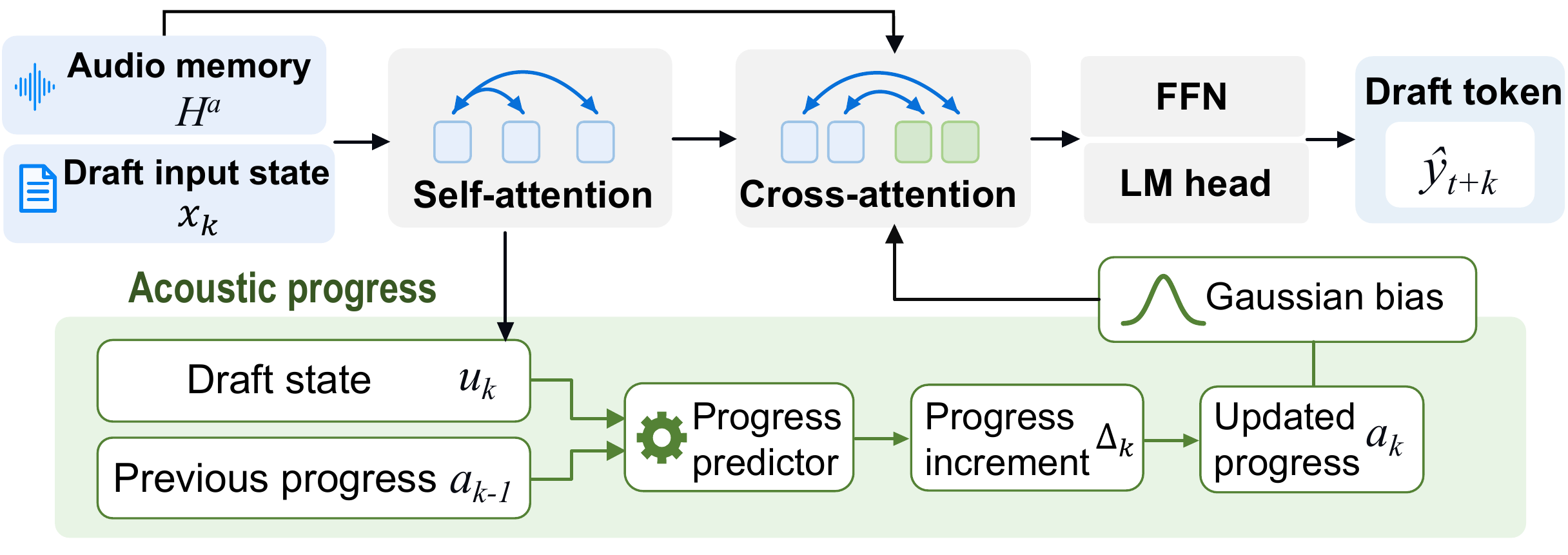}
\par\smallskip
{\small (b) Shared drafter block\par}
\end{minipage}
\caption{Overview of the proposed progress-conditioned speculative ASR decoding.
(a) Draft-and-verify pipeline. (b) Acoustic-progress prediction and feedback
within one shared draft step.}
\label{fig:drafter_architecture}
\end{figure}

\section{Method}
\label{sec:method}

\subsection{Speculative Drafter Architecture}
\label{ssec:drafter_architecture}

As illustrated in Fig.~\ref{fig:drafter_architecture}(a), given speech $X$
encoded by the target model as $H^a$ and a verified prefix
$y_{\leq t}$, the drafter proposes $K$ candidates
$\hat{\mathbf{y}}=(\hat{y}_{t+1},\ldots,\hat{y}_{t+K})$.
The target verifies them in parallel, accepting the longest prefix consistent
with its greedy predictions. Unless decoding terminates, it appends a
correction token at the first mismatch or a bonus token if all candidates
are accepted~\cite{leviathan2023speculative}.

We use a lightweight acoustic-conditioned drafter initialized by fusing the
embedding $E(y_t)$ of the last committed token with hidden features
$F_\ell(t)$ from target decoder block $\ell$ after processing that token:
\begin{equation}
x_1 = W_f\bigl[E(y_t); F_{\ell_1}(t); \cdots; F_{\ell_m}(t)\bigr],
\label{eq:drafter_initialization}
\end{equation}
where $[\,;\,]$ denotes concatenation, $W_f$ is a linear projection, and
$\ell_1,\ldots,\ell_m$ index the $m$ selected target decoder blocks.
Multi-layer target-feature fusion is also used in EAGLE-3~\cite{li2025eagle3}.
Restart features are extracted during a single-token target context update;
both this pass and batched verification are included in timing
(Sec.~\ref{ssec:experimental_setup}).

As shown in Fig.~\ref{fig:drafter_architecture}(b), a single Transformer decoder
block~\cite{vaswani2017attention} applies causal self-attention, cross-attention over
the full $H^a$, and a feed-forward network to produce draft state $z_k$ from
input $x_k$. The frozen target LM head maps $z_k$ to a token distribution,
whose argmax gives $\hat{y}_{t+k}$. The next input combines the draft state
with the new token embedding:
\begin{equation}
x_{k+1} = z_k + E(\hat{y}_{t+k}).
\label{eq:drafter_state_recurrence}
\end{equation}
The target parameters remain frozen, and the drafter hidden size matches the
target embedding and LM-head input dimensions. Draft states and the
self-attention cache are carried across steps.

\begin{table*}[t]
\centering
\caption{End-to-end speedup over target-only autoregressive decoding and average
acceptance length $\tau$ on five ASR test sets with
Qwen3-ASR-0.6B and Qwen3-ASR-1.7B. All methods use inference draft length $K=8$.}
\label{tab:main_results}
\small
\setlength{\tabcolsep}{2pt}
\renewcommand{\arraystretch}{1.12}
\begin{tabular*}{\textwidth}{@{\extracolsep{\fill}}cl*{5}{cc}@{}}
\toprule
& & \multicolumn{2}{c}{\shortstack{LibriSpeech\\test-clean}}
& \multicolumn{2}{c}{\shortstack{LibriSpeech\\test-other}}
& \multicolumn{2}{c}{\shortstack{TED-LIUM 3\\test}}
& \multicolumn{2}{c}{\shortstack{GigaSpeech\\test}}
& \multicolumn{2}{c}{\shortstack{FLEURS\\en-us test}} \\
\midrule
Model & Method & Speedup & $\tau$ & Speedup & $\tau$ & Speedup & $\tau$
& Speedup & $\tau$ & Speedup & $\tau$ \\
\midrule
\multirow{5}{*}{\shortstack{Qwen3-ASR\\0.6B}} & Base Drafter
& 1.203$\times$ & 3.043 & 1.172$\times$ & 2.976
& 1.227$\times$ & 3.102 & 1.186$\times$ & 3.012
& 1.131$\times$ & 2.864 \\
& SpeechSpec-EAGLE3
& 0.555$\times$ & 1.545 & 0.560$\times$ & 1.544
& 0.580$\times$ & 1.616 & 0.573$\times$ & 1.568
& 0.532$\times$ & 1.451 \\
& AnchorDraft
& 1.260$\times$ & 3.241 & 1.223$\times$ & 3.137
& 1.263$\times$ & 3.245 & 1.246$\times$ & 3.178
& 1.176$\times$ & 2.986 \\
& AnchorDraft + RC
& 1.120$\times$ & 3.071 & 1.105$\times$ & 3.009
& 1.179$\times$ & 3.173 & 1.120$\times$ & 3.047
& 1.065$\times$ & 2.854 \\
& \textbf{Ours}
& \textbf{1.786}$\boldsymbol{\times}$ & \textbf{5.055}
& \textbf{1.624}$\boldsymbol{\times}$ & \textbf{4.553}
& \textbf{1.737}$\boldsymbol{\times}$ & \textbf{4.867}
& \textbf{1.633}$\boldsymbol{\times}$ & \textbf{4.533}
& \textbf{1.504}$\boldsymbol{\times}$ & \textbf{4.135} \\
\midrule
\multirow{5}{*}{\shortstack{Qwen3-ASR\\1.7B}} & Base Drafter
& 0.994$\times$ & 2.871 & 0.968$\times$ & 2.781
& 1.009$\times$ & 2.918 & 0.986$\times$ & 2.839
& 0.929$\times$ & 2.663 \\
& SpeechSpec-EAGLE3
& 0.588$\times$ & 1.625 & 0.594$\times$ & 1.640
& 0.619$\times$ & 1.721 & 0.610$\times$ & 1.669
& 0.558$\times$ & 1.522 \\
& AnchorDraft
& 1.157$\times$ & 3.349 & 1.101$\times$ & 3.202
& 1.150$\times$ & 3.335 & 1.134$\times$ & 3.267
& 1.089$\times$ & 3.100 \\
& AnchorDraft + RC
& 1.059$\times$ & 3.147 & 1.021$\times$ & 3.032
& 1.090$\times$ & 3.196 & 1.057$\times$ & 3.101
& 0.997$\times$ & 2.929 \\
& \textbf{Ours}
& \textbf{1.289}$\boldsymbol{\times}$ & \textbf{4.009}
& \textbf{1.224}$\boldsymbol{\times}$ & \textbf{3.752}
& \textbf{1.291}$\boldsymbol{\times}$ & \textbf{3.967}
& \textbf{1.216}$\boldsymbol{\times}$ & \textbf{3.751}
& \textbf{1.116}$\boldsymbol{\times}$ & \textbf{3.416} \\
\bottomrule
\end{tabular*}
\end{table*}

\subsection{Acoustic Progress Propagation}
\label{ssec:acoustic_progress}

The recurrence in Sec.~\ref{ssec:drafter_architecture} propagates hidden states
and the self-attention cache but leaves acoustic progression to cross-attention.
We condition audio cross-attention on a recurrently propagated acoustic
progress state $a_k$, which estimates the temporal position in the encoded
speech at step $k$.

Following prior work~\cite{wang2026alignmentdrift}, we initialize each round
from target layer 21 (zero-indexed), using the query at the last committed
prefix token. Let $\mathcal A$ denote valid audio keys, $s_j$ their timestamps,
and $\alpha^{\mathrm{tar}}_{h,j}$ the attention weight from head $h$.
Averaging over all $N_h$ heads, we select the peak as the initial anchor:
\begin{equation}
j^\star = \arg\max_{j\in\mathcal A}
\frac{1}{N_h}\sum_{h=1}^{N_h}\alpha^{\mathrm{tar}}_{h,j},
\qquad a_1 = s_{j^\star}.
\label{eq:acoustic_anchor}
\end{equation}
These weights are reused from normal target decoding without an extra
forward pass. Later steps use drafter-predicted progress without further
target-attention queries.

For $k\ge2$, let $u_k$ denote the representation after self-attention and before
audio cross-attention. Following the monotonic position-update
principle~\cite{tjandra2017local}, we model acoustic progression as a recurrent
state transition conditioned on $u_k$ and the preceding progress state:
\begin{equation}
\begin{aligned}
\Delta_k &= \operatorname{Softplus}
\left(f_\theta\left([a_{k-1}/T;u_k]\right)\right),\\
a_k &= a_{k-1}+\Delta_k,
\end{aligned}
\label{eq:acoustic_progress_update}
\end{equation}
where $f_\theta$ is the progress predictor and $T$ is the utterance duration.
Positions and increments are measured in seconds; only the predictor's
position input is normalized by $T$.

Acoustic retrieval uses $\bar a_k$, the position clipped to $[0,T]$, to form
a Gaussian log-bias added to cross-attention scores before softmax:
\begin{equation}
\begin{aligned}
B_{k,j} &= [\Psi(\bar a_k)]_j = -\frac{(s_j-\bar a_k)^2}{2\sigma^2},\\
S'_{k,j} &= S_{k,j}+B_{k,j},
\end{aligned}
\label{eq:acoustic_progress_bias}
\end{equation}
where $S_{k,j}$ is the original score and $\sigma>0$ controls the temporal
width. This bias favors positions near $\bar a_k$ while retaining access to
the full $H^a$; recurrence and supervision use the unclipped $a_k$.

Through the recurrence in Eq.~\eqref{eq:drafter_state_recurrence}, the retrieved
acoustic context influences $u_{k+1}$ and thus the next progress prediction,
closing the feedback loop.

\subsection{Training}
\label{ssec:training}

With the target frozen, we jointly train the drafter and progress predictor
on target-generated greedy trajectories. Teacher forcing supplies
$E(y_{t+k})$ to the next step, while progress follows its own recurrence
from the target-attention anchor $a_1$, without forced-alignment resets.

We apply token cross-entropy $\mathcal L_{\mathrm{tok},i,k}$ at each unrolled
depth. Both auxiliary losses use Smooth L1 ($\beta=1$). Once per rollout,
feature matching $\mathcal L_{\mathrm{feat},i}$ compares $z_{i,1}$ with the
target's final normalized decoder state at position $t$, averaging the loss
over hidden dimensions.

We align the target-generated transcript with TorchAudio's MMS forced
aligner~\cite{pratap2024mms} and use each token's aligned time-span midpoint
as $a^*_{i,k}$. For $k\ge2$, $\mathcal L_{\mathrm{prog},i,k}$ compares the
unclipped $a_{i,k}$ with $a^*_{i,k}$ in seconds, excluding tokens without
valid alignments. Progress feedback also passes token-loss gradients to the
predictor.

For each anchor $i$, we independently sample $K_i$ and unroll
$K_i^{\mathrm{act}}=\min(K_i,R_i)$ steps, where $R_i$ counts remaining
target-trajectory tokens, including terminal EOS. Define
$m_{i,k}=\mathbf{1}\{K_i^{\mathrm{act}}\ge k\}$ and the empirical occurrence
rate $q_k=N_{\mathrm{ep}}^{-1}\sum_{i=1}^{N_{\mathrm{ep}}}m_{i,k}$ from the
epoch's rollout plan of $N_{\mathrm{ep}}$ anchors. Unexecuted steps contribute
zero loss. The per-anchor objective is
\begin{equation}
\begin{aligned}
\mathcal L_i ={}& \sum_{k=1}^{K_{\max}}
\frac{m_{i,k}\gamma^{k-1}}{q_k}\mathcal L_{\mathrm{tok},i,k}
+ \lambda_f\mathcal L_{\mathrm{feat},i} \\
&+ \lambda_p\sum_{k=2}^{K_{\max}}
\frac{m_{i,k}\gamma^{k-1}}{q_k}\mathcal L_{\mathrm{prog},i,k},
\end{aligned}
\label{eq:training_objective}
\end{equation}
where $\gamma$ controls loss decay across draft depths, and $\lambda_f$ and $\lambda_p$ weight
feature matching and progress supervision, respectively. The training
horizon range is specified in Sec.~\ref{ssec:experimental_setup}.

\section{Experiments}
\label{sec:experiments}

\subsection{Experimental Setup}
\label{ssec:experimental_setup}

\noindent\textbf{Datasets and Target Models.}
We freeze Qwen3-ASR-0.6B and Qwen3-ASR-1.7B~\cite{shi2026qwen3asr} and train
drafters on 3,933 English utterances from LibriSpeech
clean/other~\cite{panayotov2015librispeech}, TED-LIUM 3~\cite{hernandez2018tedlium3},
GigaSpeech~\cite{chen2021gigaspeech}, and FLEURS~\cite{conneau2022fleurs}.
All methods share 200 utterances per test split (1,000 total; FLEURS en-us).

\noindent\textbf{Baselines.}
\emph{Base Drafter} uses the single-layer recurrent block
(Sec.~\ref{ssec:drafter_architecture}) with unrestricted audio cross-attention
and no alignment supervision, progress prediction, or progress-based bias.
Base Drafter and \emph{AnchorDraft}~\cite{wang2026alignmentdrift} train at fixed
$K_{\mathrm{train}}=3$ unless specified. Base Drafter sums equally weighted
token losses and uses $\lambda_f=0.5$. \emph{AnchorDraft + RC} reuses the AnchorDraft
checkpoint; at inference, peak head-averaged target verification attention
centers the next round's acoustic window.
\emph{SpeechSpec-EAGLE3}~\cite{speechspec,li2025eagle3} uses native five-step
training unrolling. Drafters share each target's generated corpus.

\noindent\textbf{Metrics.}
All test token sequences matched Target-only AR, preserving word/character
error rates (WER/CER). Throughout, we report end-to-end
speedup over Target-only AR and acceptance length
$\tau=R^{-1}\sum_{r=1}^{R}(A_r+b_r)$, pooling all $R$ rounds per evaluated
set. $A_r$ counts accepted draft-prefix tokens and $b_r\in\{0,1\}$ the emitted
correction/bonus token ($b_r=0$ if EOS terminates without one).

\noindent\textbf{Implementation.}
Except for the frozen-drafter ablation, drafters train for 45 epochs
(14,760 updates) at effective batch size 12 for both targets.
Our drafters contain 17.85M/71.34M parameters (0.6B/1.7B targets)
and use features from decoder blocks 0, 9, 18, and 27 (zero-indexed).
Our joint training uses AdamW~\cite{loshchilov2019adamw} (learning rate $10^{-3}$,
cosine decay) and independent per-anchor sampling
$K\sim\mathcal U\{3,\ldots,8\}$ (Sec.~\ref{ssec:training}).
We set $\sigma=0.2$\,s, $\gamma=0.7$, $\lambda_f=0.5$, and $\lambda_p=0.1$.
All methods decode greedily; end-to-end timing uses one H100 80GB GPU,
batch size 1, FP32 eager attention for the target, and BF16 autocast for the drafter.

\subsection{Main Results}
\label{ssec:main_results}

Across all five test sets and both target model sizes, our method achieved
the highest end-to-end speedup and average acceptance length
(Table~\ref{tab:main_results}). Relative to AnchorDraft, the macro-averaged speedup and
$\tau$ improved by 34.3\% and 46.6\%, respectively, with Qwen3-ASR-0.6B,
and by 9.0\% and 16.3\% with Qwen3-ASR-1.7B. AnchorDraft consistently
outperformed Base Drafter, but adding runtime correction reduced both
metrics. SpeechSpec-EAGLE3 did not achieve end-to-end speedup under the
shared-training-corpus setting.

\begin{figure}[t]
\centering
\definecolor{HorizonBase}{HTML}{777777}
\definecolor{HorizonAnchor}{HTML}{E69F00}
\definecolor{HorizonRC}{HTML}{0072B2}
\definecolor{HorizonRandom}{HTML}{CC79A7}
\definecolor{HorizonRandomRC}{HTML}{D55E00}
\definecolor{HorizonOurs}{HTML}{009E73}
\pgfplotslegendfromname{draft_horizon_legend}
\par\smallskip
\begin{tikzpicture}
\begin{groupplot}[
    group style={group size=2 by 1, horizontal sep=0.165\columnwidth},
    scale only axis,
    width=0.34\columnwidth,
    height=0.27\columnwidth,
    axis lines=box,
    axis line style={black!75, thin},
    xmin=2.6, xmax=12.4,
    xtick={3,4,6,8,10,12},
    xlabel={Draft horizon $K$},
    tick align=inside,
    tick pos=left,
    tick label style={font=\small},
    scaled y ticks=false,
    yticklabel style={/pgf/number format/fixed,
        /pgf/number format/precision=1, /pgf/number format/fixed zerofill},
    label style={font=\small},
    ymajorgrids=false,
    xmajorgrids=false,
    every axis plot/.append style={line width=0.8pt, mark size=1.5pt,
        mark options={solid}},
    legend style={font=\small, draw=none,
        /tikz/every even column/.append style={column sep=7pt}},
]
\nextgroupplot[
    ylabel={Acceptance length $\tau$},
    ymin=2.5, ymax=5.2,
    ytick={2.5,3.0,3.5,4.0,4.5,5.0},
    legend to name=draft_horizon_legend, legend columns=3, legend transposed=true,
]
\addplot[color=HorizonBase, solid, mark=*] coordinates {
    (3,2.9249) (4,2.9927) (6,2.9977) (8,2.9971) (10,2.9964) (12,2.9974)
};
\addlegendentry{Base Drafter}
\addplot[color=HorizonAnchor, dashdotted, mark=triangle*] coordinates {
    (3,3.0223) (4,3.1374) (6,3.1530) (8,3.1527) (10,3.1523) (12,3.1523)
};
\addlegendentry{AnchorDraft}
\addplot[color=HorizonRC, dashed, mark=square*] coordinates {
    (3,2.9561) (4,3.0226) (6,3.0267) (8,3.0271) (10,3.0274) (12,3.0271)
};
\addlegendentry{AnchorDraft + RC}
\addplot[color=HorizonRandom, densely dotted, mark=pentagon*] coordinates {
    (3,2.9169) (4,3.0768) (6,3.1568) (8,3.1669) (10,3.1696) (12,3.1692)
};
\addlegendentry{AnchorDraft + VHT}
\addplot[color=HorizonRandomRC, densely dashdotted, mark=x] coordinates {
    (3,2.8393) (4,2.9278) (6,2.9440) (8,2.9440) (10,2.9437) (12,2.9430)
};
\addlegendentry{AnchorDraft + VHT + RC}
\addplot[color=HorizonOurs, solid, mark=diamond*] coordinates {
    (3,3.2477) (4,3.6841) (6,4.2661) (8,4.6011) (10,4.7857) (12,4.8940)
};
\addlegendentry{Ours}
\nextgroupplot[
    ylabel={Speedup ($\times$)},
    ymin=0.95, ymax=1.8,
    ytick={1.0,1.2,1.4,1.6,1.8},
]
\addplot[color=HorizonBase, solid, mark=*] coordinates {
    (3,1.2890) (4,1.2855) (6,1.2461) (8,1.1829) (10,1.1248) (12,1.0778)
};
\addplot[color=HorizonAnchor, dashdotted, mark=triangle*] coordinates {
    (3,1.3253) (4,1.3472) (6,1.2938) (8,1.2323) (10,1.1739) (12,1.1134)
};
\addplot[color=HorizonRC, dashed, mark=square*] coordinates {
    (3,1.2666) (4,1.2549) (6,1.1854) (8,1.1172) (10,1.0580) (12,1.0031)
};
\addplot[color=HorizonRandom, densely dotted, mark=pentagon*] coordinates {
    (3,1.2862) (4,1.3223) (6,1.3004) (8,1.2448) (10,1.1912) (12,1.1306)
};
\addplot[color=HorizonRandomRC, densely dashdotted, mark=x] coordinates {
    (3,1.2090) (4,1.2156) (6,1.1525) (8,1.0940) (10,1.0249) (12,0.9768)
};
\addplot[color=HorizonOurs, solid, mark=diamond*] coordinates {
    (3,1.3719) (4,1.4901) (6,1.6268) (8,1.6487) (10,1.5792) (12,1.5224)
};
\end{groupplot}
\end{tikzpicture}
\caption{Average acceptance length $\tau$ (left) and end-to-end speedup over
Target-only AR (right) versus inference horizon $K$, with Qwen3-ASR-0.6B on
the 1,000-utterance test set.}
\label{fig:draft_horizon}
\end{figure}
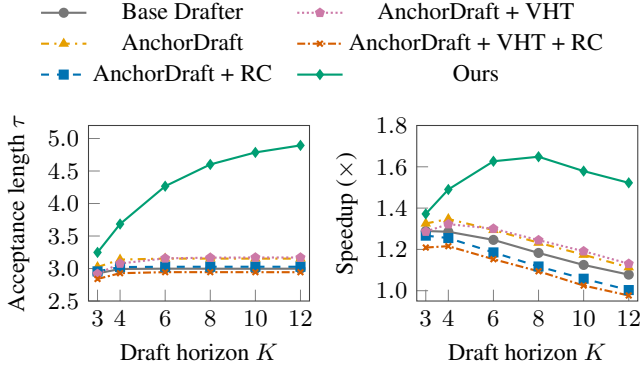

Our drafter's acceptance length continued to grow beyond baseline
saturation (Fig.~\ref{fig:draft_horizon}). Because verification accepts only
a consecutive matching prefix, early mismatches limit the benefit of
extending a draft. AnchorDraft with variable-horizon training
(VHT; $K\sim\mathcal U\{3,\ldots,8\}$) plateaued by $K=6$, with or without
runtime correction, supporting acoustic
progress propagation under matched training horizon ranges. Our method
achieved the highest speedup at every tested horizon, peaking at $K=8$.
Beyond this point,
further acceptance gains did not offset the increased cost of longer
draft-and-verify rounds.

\subsection{Ablation Study}
\label{ssec:ablation_study}

We evaluate six ablations (Table~\ref{tab:ablation}).
\emph{w/o Acoustic Progress} matches Base Drafter's architecture and inference,
training from scratch with our variable horizons and reweighted token loss
(Eq.~\eqref{eq:training_objective}).
\emph{w/o Progress Feedback} removes attention feedback but retains progress
prediction and supervision. \emph{Absolute-position prediction} uses
$a_k=g_\phi(u_k)$ without position recurrence, retaining supervision and feedback.
\emph{w/o Joint Training} freezes the final Base Drafter. A randomly initialized
predictor trains for five position-only epochs (selecting epoch 3), then
five variable-horizon epochs with token CE and progress loss ($\lambda_p=0.1$).
We report the final checkpoint.
\emph{Fixed-$K$ training} fixes $K=3$ or $K=8$.

\begin{table}[!htbp]
\centering
\caption{Ablations on Qwen3-ASR-0.6B using 1,000 test utterances
and inference horizon $K=8$.}
\label{tab:ablation}
\small
\setlength{\tabcolsep}{3pt}
\renewcommand{\arraystretch}{1.12}
\begin{tabular*}{\columnwidth}{@{\extracolsep{\fill}}lcc@{}}
\toprule
Model Variant & \shortstack{Acceptance\\length $\tau\uparrow$}
& \shortstack{E2E\\Speedup $\uparrow$} \\
\midrule
w/o Acoustic Progress & 3.078 & 1.206$\times$ \\
w/o Progress Feedback & 3.069 & 1.210$\times$ \\
Absolute-position prediction & 3.214 & 1.233$\times$ \\
w/o Joint Training    & 3.130 & 1.121$\times$ \\
Fixed-$K=3$ training & 3.760 & 1.331$\times$ \\
Fixed-$K=8$ training & 4.572 & 1.611$\times$ \\
\midrule
\textbf{Ours} & \textbf{4.601} & \textbf{1.649}$\boldsymbol{\times}$ \\
\bottomrule
\end{tabular*}
\end{table}

Table~\ref{tab:ablation} supports the contribution of cross-step propagation:
recurrence outperformed absolute-position prediction in both metrics with
supervision and feedback retained, whereas supervision without feedback
performed comparably to removing progress modeling. An explicit position
reference may ease tracking compared with recovering absolute location from
each hidden state.

The frozen-drafter variant underperformed joint training, but the schedules
differ, so this comparison does not isolate joint optimization.
Variable-horizon training outperformed fixed-$K=8$ training in both metrics
at the same maximum horizon, indicating gains beyond exposure to longer
drafts. The fixed-horizon model also improved both metrics over $K=3$.

\section{Conclusion}
\label{sec:conclusion}

We have proposed a speculative ASR drafter that propagates acoustic progress
to guide audio cross-attention. Joint training over variable horizons
improves its use of longer drafts. Across five ASR test sets and two target
sizes, higher acceptance lengths and lossless end-to-end speedups over the
evaluated baselines support acoustic progress propagation beyond local
alignment alone.

\FloatBarrier
\clearpage
\bibliographystyle{IEEEbib}
\bibliography{strings,refs}

\end{document}